# The State of Speech in HCI: Trends, Themes and Challenges

Leigh Clark[1], Phillip Doyle[1], Diego Garaialde[1], Emer Gilmartin[2], Stephan Schlögl[3], Jens Edlund[4], Matthew Aylett[5], João Cabral[6], Cosmin Munteanu[7], Benjamin Cowan[1]

[1]School of Information and Communication Studies, University College Dublin, Ireland
[2]Speech Communication Laboratory, Trinity College Dublin, Ireland
[3]MCI Management Center, Innsbruck, Austria
[4]Department of Speech, Music and Hearing, KTH Royal Institute of Technology, Sweden
[5]School of Informatics, University of Edinburgh, United Kingdom
[6]School of Computer Science and Statistics, Trinity College Dublin, Ireland
[7]Institute of Communication, Culture, Information and Technology, University of Toronto Mississauga, Canada

**Abstract**
Speech interfaces are growing in popularity. Through a review of 68 research papers this work maps the trends, themes, findings and methods of empirical research on speech interfaces in HCI. We find that most studies are usability/theory-focused or explore wider system experiences, evaluating Wizard of Oz, prototypes, or developed systems by using self-report questionnaires to measure concepts like usability and user attitudes. A thematic analysis of the research found that speech HCI work focuses on nine key topics: system speech production, modality comparison, user speech production, assistive technology & accessibility, design insight, experiences with interactive voice response (IVR) systems, using speech technology for development, people's experiences with intelligent personal assistants (IPAs) and how user memory affects speech interface interaction. From these insights we identify gaps and challenges in speech research, notably the need to develop theories of speech interface interaction, grow critical mass in this domain, increase design work, and expand research from single to multiple user interaction contexts so as to reflect current use contexts. We also highlight the need to improve measure reliability, validity and consistency, in the wild deployment and reduce barriers to building fully functional speech interfaces for research.

**Author Keywords**
Speech interfaces; speech HCI; review; speech technology; voice user interfaces

**Research Highlights**
- Most papers focused on usability/theory-based or wider system experience research with a focus on Wizard of Oz and developed systems, though a lack of design work
- Questionnaires on usability and user attitudes often used but few were reliable or validated
- Thematic analysis showed nine primary research topics
- Gaps in research critical mass, speech HCI theories, and multiple user contexts

Email: leigh.clark@ucd.ie; phillip.doyle1@ucdconnect.ie; diego.garaialde@ucdconnect.ie; gilmare@tcd.ie; stephan.schloegl@mci.edu; edlund@speech.kth.se; matthewaylett@gmail.com; cabralj@scss.tcd.ie; cosmin@taglab.ca; benjamin.cowan@ucd.ie.

# 1. Introduction

Speech has become a more prominent way of interacting with automatic systems. In addition to long established telephony based or interactive voice response (IVR) interfaces, voice enabled intelligent personal assistants (IPAs) like Amazon Alexa, Apple Siri, Google Assistant, and Microsoft Cortana are widely available on a number of devices. Home-based devices such as Amazon Echo, Apple HomePod, and Google Home are increasingly using speech as the primary form of interaction. The market for IPAs alone is projected to reach $4.61 billion by the early 2020s (Kamitis, 2016). The technical infrastructures underpinning speech interfaces have advanced rapidly in recent years and is the subject of extensive research in the speech technology community (Chan, Jaitly, Le, & Vinyals, 2016). Research on the user side of speech interfaces is said to be limited by comparison (Aylett, Kristensson, Whittaker, & Vazquez-Alvarez, 2014) yet we currently know little about the current state of this research in the human-computer interaction (HCI) field. Our work aims to map out the current state of speech interface work in HCI, giving a comprehensive review of empirical research published in core HCI sources on speech interface interaction. We particularly concentrate on uncovering the methods and approaches used for studying speech interaction in HCI and the topics covered in this research.

Our review finds that the majority of research in this domain is usability/theory-focused or explores wider system experiences. Quantitative methodological approaches are particularly dominant, with relatively few qualitative or design-oriented studies. The stimuli and measures used to research speech also vary. Research uses a variety of Wizard of Oz (WoZ) systems, interactive prototypes, or established commercial systems. Self-report questionnaires, measuring concepts like usability and user attitudes towards interaction, are also common, yet the use of standard validated questionnaires is low. Most studies use bespoke measures that are not adequately tested for reliability or validity. The research conducted also tends to converge on nine topics: system speech production, modality comparison, user speech production, assistive technology & accessibility, design insight, experiences with IVR systems, using speech technology for development, people's experiences with IPAs and how user memory affects speech interface interaction. Based on our review, we propose a number of research challenges for the HCI field to address in future speech-based HCI research. These include the need to 1) devise theories of speech interface interaction; 2) strive towards achieving a critical mass of research in multiple areas of speech interaction; 3) grow speech interaction design work and 4) research user experience (UX) issues across conversational contexts with multiple users. We also propose a number of evaluation challenges, in particular 1) the need to improve the reliability, validity and consistency of evaluation measures; 2) increase the level of in the wild evaluation, and 3) making the creation of deployable speech interface prototypes easier.

## 2. What are speech interfaces?

Speech interfaces are systems that use speech, either pre-recorded ('canned') or synthesised speech to communicate or interact with users (Weinschenk & Barker, 2000). Speech can be used as a primary input or output, or in a two-way dialogue. Systems using speech (canned or synthesised) as their primary medium for communicating information to the user (Aylett, Vazquez-Alvarez, & Baillie, 2015) are analogous to audio user interfaces (Weinschenk & Barker, 2000). In this type of interaction speech is monologic; produced only by the system as output. Speech can also be used solely as an input modality, often seen in 'command and control' of systems or devices (Cohen et al., 2013). In this, user input is recognised by *automatic speech recognition* (ASR) and *natural language understanding* (NLU) components to identify user intents or commands. Perhaps the most common example of speech interfaces are spoken dialogue systems (SDS), where the system and user can interact through spoken natural language. The dialogue involved can range from highly formulaic question-answer pairs where the system keeps the initiative and users respond, through HMIHY ('How may I help you?') systems where the user can formulate wider queries and the system computes the optimal next move, to systems which appear to allow user initiative by permitting users to interrupt (*'barge-in'*) and change task. However, at present, system dialogs are quite stiff and just how closely they simulate human conversation is open to debate. SDSs generally follow a common pipeline design to engage in dialogue with users. They first detect that a user is addressing the system. This can include identifying the user addressing it from a range of possible users (speaker diarization). The system's ASR then recognizes what is said and passes recognition results to the NLU component. The function of the NLU is to identify the intent behind the user's utterance, and express it in machine-understandable form, often as a *dialog act*. The system *dialog manager* (DM) then selects an appropriate action to take based on the identified intent, considering factors such as the current state of the dialogue and recent dialog history. A *natural language generation* (NLG) component then generates a natural response, which is outputted by the system as artificial speech using *text-to-speech synthesis* (TTS) (Jokinen & MacTear, 2010; Lison & Kennington, 2016). Some speech interfaces are used in conjunction with other input/output modalities to facilitate interaction (Weinschenk & Barker, 2000). For example, speech can be used as either input or output along with graphical user interfaces (GUIs), commonly seen in speech dictation for word-processing tasks, or in screen reading technology to support the navigation of websites. Indeed, SDS's such as IPAs (e.g. Siri) often display content and request user input through the screen as well as through speech (Cowan et al., 2017).

## 2.1. Research Aims

While interest in speech interfaces has been growing steadily (Cohen, Cheyer, Horvitz, El Kaliouby, & Whittaker, 2016; Munteanu et al., 2017; Munteanu & Penn, 2014) there is no clear idea of what forms the core of speech-based work in HCI. This makes it difficult to identify novel areas of research and the challenges faced in the HCI field, particularly for those new to the topic. As speech gains in popularity as an interface modality, it is important that the state of speech research published in the HCI community is clearly mapped, so that those who come to the research have a clear idea of the major trends, topics and methods. The current paper aims to achieve this by reviewing empirical work published across a range of leading HCI venues. We also hope this may guide and inform future endeavours in the field, by identifying opportunities for further research. Below we report the method used to conduct the review, our findings, and discuss challenges for future research efforts based on our results.

## 3. Method

### 3.1. Scope

We reviewed 68 publications on user interactions with speech as either a system output (e.g. Alm, Todman, Elder, & Newell, 1993), user input (e.g. Harada, Wobbrock, Malkin, Bilmes, & Landay, 2009) or in a dialogue context (e.g. Cowan, Branigan, Obregón, Bugis, & Beale, 2015; Porcheron, Fischer, & Sharples, 2017). Papers were selected using adapted PRISMA guidelines, similar to the adapted QUORUM procedures in previous reviews (Bargas-Avila & Hornbæk, 2011; Mekler, Bopp, Tuch, & Opwis, 2014).

## 3.2. Search procedure

Three databases were searched for relevant publications in January 2018: ACM Digital Library (ACM DL), ProQuest (PQ) and Scopus (SP). Each database was searched using terms generated from keywords in existing speech literature, and from a survey of 11 leading researchers in speech HCI and speech technology (see Table 1). The terms were searched as exact phrases and, where possible, combined using Boolean operators (e.g. "OR"). Otherwise, terms were searched for individually. Searches were limited to terms appearing in the title, abstract, and publication keywords, and were also limited to journal articles and conference papers, excluding other sources such as technical reports (see 3.3 for further details). Search results were limited to core HCI publication venues, which were established by combining top HCI publication sources listed on Google Scholar, Thomson Reuters, and Scimago journal rankings. Duplicates were removed, leaving 48 unique HCI publication venues. This list is available in the supplementary material. Searches were then imported into the reference management software Mendeley.

> speech interface; voice user interface; voice system; human computer dialog*; human machine dialog*; natural language dialog* system; natural language interface; conversational interface; conversational agent; conversational system; conversational dialog* system; automated dialog* system; interactive voice response system; spoken dialog* system; spoken human machine interaction; human system dialog*; intelligent personal assistant

**Table 1. Review search terms. Asterisks (*) denote truncation to account for alternative spellings e.g. *dialog* or *dialogue*.**

## 3.3. Inclusion and exclusion criteria

The search resulted in a total of 1,181 unique entries, following the removal of duplicates (ACM DL=722, PQ=83, SP=376). Several inclusion and exclusion criteria were then applied to ensure relevance to the aims of the research. Inclusions were based on the following criteria: (1) *Only papers primarily investigating speech input, speech output and/or dialogue were included*. Only papers that investigated speech as a user input (user utterances to system), output (output from system to user) or user-system dialogue (two-way interaction) were included in the final set of papers. (2) *Only full papers that were published in conferences and journals and written in English were included*. Because of the similarity in status with full papers, CHI notes were also included.

Exclusions were based on the following criteria: (1) *Papers investigating embodied interfaces were excluded*. Papers that discussed embodied interfaces like embodied conversational agents (e.g. Bickmore, Pfeifer, & Jack, 2009) and robots (e.g. Strait, Vujovic, Floerke, Scheutz, & Urry, 2015) were removed. This was to avoid studies where speech is confounded by issues of embodiment, such as gesture and emotive facial expressions (Breazeal, 2003), that can affect users' interactions and experiences (Bruce, Nourbakhsh, & Simmons, 2002; Kuno et al., 2007). (2) *Papers without empirical measurement or evaluation of interaction with users were excluded*. We excluded technical discussions of systems, models or methods related to speech interfaces that had little to no user evaluation (Han, Philipose, & Ju, 2013). *(3) Non-full or non-peer reviewed papers were excluded*. We excluded works in progress and extended abstracts (e.g. Aylett et al., 2015; Cowan et al., 2016), debates or panel discussions (e.g. Cohen et al., 2016) workshop papers (e.g. Munteanu et al., 2017) and magazine articles, for instance those from Communications of the ACM (e.g. Shneiderman, 2000) or ACM Interactions (e.g. Shneiderman & Maes, 1997).

The lead author and a co-author filtered the results independently based on these selection criteria. There was a strong level of agreement between the authors in the final set of papers chosen [Kappa = 0.843 ($p < .001$), 95% CI (0.729, 0.957)]. After applying these criteria, 68 papers were selected for final analysis. Following Mekler et al. (2014) we extracted the research aims, interaction type being researched, methodologies, and results from each paper. We also extracted: publication sources; research topics; types of interfaces used; aspects of speech interfaces being assessed; types of research conducted; experiment procedures and measures, and participant demographics. This information is available in the supplementary material.

## 4. Results: Publication Trends & Research Methodologies

### 4.1. Publication Trends

There was an observed skew towards speech HCI work appearing in conferences rather than journals, with 63.2% of papers (N=43) published in conferences, and 36.8% (N=25) published in journals. While there were 14 unique publication venues overall, CHI was by far the most common venue for work reviewed here, accounting for 39.7% (N=27) of all papers. The International Journal of Human-Computer Studies was the most common journal publication to feature, accounting for 14.7% of papers reviewed (N=10).

| Publication | Count |
| --- | --- |
| Conference on Human Factors in Computing Systems (CHI)[1] | 27 |
| International Journal of Human-Computer Studies (IJHCS)[2] | 10 |
| International Conference on Multimodal Interaction (ICMI) | 7 |
| Behaviour & Information Technology | 5 |
| ACM Transactions on Computer-Human Interaction (TOCHI) | 3 |
| Interacting with Computers | 3 |
| Conference on Human-Computer Interaction with Mobile Devices and Services (Mobile HCI) | 3 |
| Computers in Human Behavior | 2 |
| Computer-Supported Cooperative Work and Social Computing (CSCW) | 2 |
| Intelligent User Interfaces (IUI) | 2 |
| ACM Transactions on Interactive Intelligent Systems (TIIS) | 1 |
| Designing Interactive Systems (DIS) | 1 |
| IEEE International Conference on Systems, Man and Cybernetics | 1 |
| User Modeling and User-Adapted Interaction | 1 |

**Table 2. Number of papers reviewed by publication source.**

---

[1] This includes one publication at the INTERCHI '93 conference - a joint conference between INTERACT and CHI.
[2] Includes papers published under the journal's previous title "International Journal of Man-Machine Studies".

### 4.2. Research Methods

This section provides a summary of the methodological approaches used in the 68 papers reviewed, highlighting the systems used in interactions, communicative contexts, and measures to research speech in HCI (see Table 3). Some papers contained more than one type of system in the categories presented (e.g. Medhi, Gautama, & Toyama, 2009) and as such the multiple systems are included in the totals.

| Device contexts | Mock systems | | | Existing systems & working prototypes | | |
|---|---|---|---|---|---|---|
| **Type of Speech Speech Interface** | *User speech input* | *System speech output* | *User-system dialogue* | *User speech input* | *System speech output* | *User-system dialogue* |
| Computer-based | 6 | 2 | 7 | 8 | 8 | 3 |
| Telephone-based | - | - | 8 | - | 1 | 15 |
| Mobile applications/IPAs | - | - | 1 | 1 | - | 4 |
| Vehicle-based | - | - | 2 | - | - | 2 |
| Other | - | - | - | 1 | 1 | 1 |

**Table 3: Frequency of systems tested.**

*4.2.1. Systems tested*

One third of systems in the research (N=26) were *mock systems*, in which users interacted with a Wizard of Oz system. In WoZ studies participants are led to believe they are interacting with a system, however, a (usually unseen) confederate controls the system's output (Dahlbäck, Jönsson, & Ahrenberg, 1993). These studies are generally designed to assess user responses to different types of output from hypothetically possible systems (e.g. Knutsen, Le Bigot, & Ros, 2017; Oviatt, Swindells, & Arthur, 2008). 45 of the papers reviewed either focused on *existing systems* or *working prototypes*. These included existing IPAs like Amazon Alexa and Siri (e.g. Cowan et al., 2017; Porcheron et al., 2017), prototype speech systems (e.g. Yankelovich, Levow, & Marx, 1995), and custom-built interfaces for delivering user-directed speech (e.g. Clark, Ofemile, Adolphs, & Rodden, 2016). One system reported in the papers, but excluded from Table 3, was a *hypothetical interaction* - one with a more general focus on computers that use speech as a whole (Buchheit & Moher, 1990) - rather than exploring a specific type of system.

*4.2.2. Device contexts*

We discovered four main types of device contexts that researchers used to deliver speech interfaces. 48.6% (N=35) researched *computer-based interactions*: existing or customized software running on desktops or laptops to accomplish a range of specific tasks. In these interactions, users sat with a computer and interacted with the speech interface for the duration of the experiment. Tasks included completing picture-naming tasks with multiple partner conditions (Cowan et al., 2015) or listening to computer-media news (Kallinen & Ravaja, 2005). 33.3% (N=24) of systems explored *telephone-based dialogues* as an interaction, for example in healthcare appointment booking (Wolters et al., 2009) and telephone banking systems (e.g. Wilkie, Jack, & Littlewood, 2005). Six speech systems were *mobile applications* or *IPAs* on smartphones, including existing mobile IPAs (e.g. Porcheron et al., 2017), and a mobile interface using speech input (Corbett & Weber, 2016). Four of the systems were *vehicle-based* systems, where studies examined the user responses to system speech within driving simulation tasks (e.g. Truschin, Schermann, Goswami, & Krcmar, 2014). Finally, three were classified as *other types* of device contexts that did not readily fit into the preceding categories. The first of these papers evaluated a computer output model based on human tutor feedback (Porayska-Pomsta & Mellish, 2013), while the second examined the use of a hands-free audio device with a speech interface in a healthcare setting (Sammon, Brotman, Peebles, & Seligmann, 2006). Finally, Piper & Hollan (2008) explored the use of tabletop displays to support medical conversations between deaf and hearing individuals, allowing a medical doctor to provide spoken input that was converted into text.

*4.2.3. Direction of Communication*

We also categorized the direction of communication being explored in each paper, showing the breakdown of the papers across our three types of speech interface outlined in the introduction: *user speech input* (monologic from user to system), *system speech output* (monologic from system to user), and *user-system dialogue* (two-way speech interaction). Of the monologic interactions, 12 systems explored only system speech output. For example, asking users to evaluate voices (Dahlbäck, Wang, Nass, & Alwin, 2007) or assess an auditory web-browser (Sato, Zhu, Kobayashi, Takagi, & Asakawa, 2011). 16 systems used only user speech input, including the use of an alternative interaction model for dictation tasks (Kumar, Paek, & Lee, 2012), or in assessing elderly users' interactions with an IVR system (Murata & Takahashi, 2002). The overwhelming majority of systems investigated some form of dialogue between system and user, in which varying degrees of two-way spoken communication was explored (N=44). Examples included assessing language production when interacting with mocked up speech interfaces (e.g. Amalberti, Carbonell, & Falzon, 1993; Cowan et al., 2015), interactions with existing commercially available IPAs (e.g. Luger & Sellen, 2016; Porcheron et al., 2017) and telephone-based IVR systems (e.g. Wilke, McInnes, Jack, & Littlewood, 2007; Wilkie et al., 2005).

*4.2.4. Measures*

In measuring user interactions with systems, the majority of papers (N=37) used a combination of objective (e.g. user speech choices, task completion time) and subjective (e.g. self-report questionnaire) measures. Fewer papers used only subjective (N=19), while 12 papers relied solely on objective measures.

A number of concepts were measured across the research reviewed (see Table 4). *User attitudes* were most commonly measured. These included attitudes towards an interface's voice and/or speech content with research measuring concepts such as likeability and human likeness (e.g. Clark et al., 2016). *Task performance* measures like the total number of turns (Le Bigot et al., 2007) percentage of tasks completed correctly (Oviatt et al., 2008) and task completion time (Patel et al., 2009) were also commonly used. *Lexis and syntax* choices were also measured in the research reviewed (Cowan et al., 2015; Piper & Hollan, 2008). *Perceived usability* measures concentrated on quantifying concepts like perceived ease of use and learnability (e.g. Evans & Kortum, 2010), through Likert scale questionnaires. *System usage* measures aimed to identify and quantify what people used speech interfaces for (e.g. Cowan et al., 2017) and how they used them (e.g. Schaffer, Schleicher, & Möller, 2015). Memory based research included measures of *user recall* in specific aspects of interaction (e.g. recall of system outputs (Knutsen et al., 2017)). *Physiological* data was assessed with measure including eye tracking (Hofmann, Tobisch, Ehrlich, Berton, & Mahr, 2014), or loudness of speech (Lunsford, Oviatt, & Coulston, 2005). *Other* measures included system recognition of user input (e.g. Oviatt et al., 2008), user agency (Limerick, Moore, & Coyle, 2015), and assessing user creativity and self-disclosure (Wang & Nass, 2005), and cognitive load (Truschin et al., 2014).

| Concepts measured | Objective | Subjective | Total |
|---|---|---|---|
| User attitudes | - | 36 | 36 |
| Task performance | 28 | 5 | 33 |
| Lexis & syntax | 19 | 1 | 20 |
| Perceived usability | - | 18 | 18 |
| System usage | 13 | 12 | 15 |
| User recall | 6 | 1 | 7 |
| Physiological data | 3 | - | 3 |
| Other | 2 | 9 | 11 |

**Table 4. Frequency of objective and subjective measurement of concepts.**

*4.2.5. Data collection methods*

Many papers included a combination of measures in conducting their research. *Questionnaires* were used in the majority of papers (N=40), assessing aspects such as user attitudes (e.g. Suhm et al., 2002) and usability (e.g. Perugini, Anderson, & Moroney, 2007). These tended to be administered post interaction. The majority of these scales were *custom-built* (N=38) and varied from single to multiple-item questionnaires. Some of these custom-built scales were based on pre-existing measures. For instance, Le Bigot et al. (2006) developed a set of custom items based on the NASA-TLX scale for measuring cognitive load. Similarly, Wolters et al. (2009) formed a scale based on the ITU-T Rec. P.815 evaluation for telephone-based dialogue systems (Möller, Engelbrecht, Kuhnel, Wechsung, & Weiss, 2009). Usability measures were occasionally assessed using existing scales, with 3 papers using the System Usability Scale (SUS) (e.g. Evans & Kortum, 2010). Three other existing scales were used in the papers reviewed. The unmodified NASA-TLX scale for measuring cognitive load appeared three times (e.g. Kumar et al., 2012). The Subjective Assessment of Speech System Interface (SASSI) and the Driving Activity Load Index (DALI) were both used once within the same paper (Hofmann et al., 2014). The *Observations* of participants' interaction behaviour were also common (N=36). Examples include lexical choices (e.g. Porcheron et al., 2017) and modality use (e.g. (e.g. Melichar & Cenek, 2006). *Interviews* were less common (N=20), and mainly focused on participants reporting and reflecting on their interaction experiences (e.g. Luger & Sellen, 2016). *System measures* like system logs appeared in fewer papers (N=18) and were useful for measuring what people used systems for (Sammon et al., 2006). *Other* data collection methods, which did not fit easily into the other categories, appeared five times in the papers. These included corpus data (Derriks & Willems, 1998), focus groups (Cowan et al., 2017), user modelling (Schaffer et al., 2015), perceived sense of agency (Limerick et al., 2015) and physiological data (Kallinen & Ravaja, 2005). One paper was not explicitly clear in the methods for evaluating the iterative design of a tutor system, though it is assumed some form of interview was used (Hakulinen, Turunen, Salonen, & Räihä, 2004). The frequency of each data collection method is included in Table 5.

| Data collection methods | Number of papers |
|---|---|
| Questionnaire | 40 |
| Observations | 36 |
| Interview | 20 |
| System measures | 18 |
| Other | 5 |
| Not explicit | 1 |

**Table 5. Frequency of data collection methods.**

## 5. Results: Research Topics

As part of our review we also categorised the types of work reviewed (5.1) as well as the primary research themes covered across the papers (5.2).

### 5.1. Type of work

The papers reviewed generally divided into two types of work. The majority of papers (N=43) were categorized as *usability/theory-based research*. These papers comprised of: (a) those exploring how a particular design choice or behaviour impacted usability, systems, or user performance and/or UX measures; and (b) those exploring concepts and theories from research in human communication (e.g. linguistics, psychology) in the context of speech interface interaction. These two approaches often overlapped, and a number of papers in this category used theory or theory-based concepts to inspire the design of specific aspects of a speech interface, rather than researching complete systems. For example, Wilkie et al. (2005) used politeness theory (Brown & Levinson, 1987) to design greetings with specific politeness strategies in voice enabled phone banking systems and assessed usability and user attitudes towards the system based on these strategies. Similarly, other papers in this category have modified specific components, such as synthesised speech, while also exploring theories borrowed from human communication (e.g. alignment (Cowan et al., 2015) and vague language (Clark et al., 2016)).

The remaining papers (N=25) were classified as *system experience research*. Unlike the previous category, these papers explored user interactions with either working prototypes or existing systems, rather than specific system components and did not focus on theory-based concepts. This incorporated the vast majority of exploratory work that used semi-structured interviews, ethnography, and other qualitative analysis techniques to identify user views and issues. For example, recent work by Luger & Sellen (2016) and Cowan et al. (2017) explored users' experiences and past interactions with IPAs through interviews and focus groups respectively. A number of system focused research papers explored the deployment of bespoke and prototypical systems. Corbett & Weber (2016), for example, conducted usability studies of a system designed specifically for motor-impaired users. Similar approaches were adopted in exploring speech interface systems for users with other specific accessibility requirements (e.g. Harada et al., 2009; Piper & Hollan, 2008), healthcare workers (Lai & Vergo, 1997; Sammon et al., 2006) and identifying more general challenges towards designing an experimental system (Yankelovich et al., 1995).

### 5.2. Research Themes

Inductive Thematic Analysis (Braun & Clarke, 2006) was conducted on the final search results to categorize the research themes discussed in each publication. Themes were initially coded independently by two of the authors. After initial coding, perceived inconsistencies and variation in themes were resolved by discussion between the two authors and an independent observer, who had familiarity with the papers but had not contributed to the initial coding. Table 6 shows the breakdown of papers for each of the research topics and, if appropriate, their respective sub-topics. In some cases, the focus of papers overlapped topics and sub-topics, and so papers are categorised by their primary research topic, which was judged collaboratively to be the main focus of the paper. The topics and findings for the 68 papers are summarised in the sections below.

*5.2.1. System speech production*

26.5% (N=18) of the papers reviewed focused on the topic of *system speech production* – that is, system speech directed towards its users – and its effects on interaction behaviour and/or UX. When discussing this in the papers, speech interfaces tend to be referred to as the *computer, machine, interface* or *system*. This was the largest category observed, and was organised into a further four sub-topics, each focusing on a more specific area of system speech production.

The first of these sub-topics contains papers discussing elements of *synthesis* in a system's speech production. In these papers, authors investigated the effects of manipulating aspects of a system's voice on users' experience and performance. Two papers examined the effects of synthesised speech on perceived personality. In the context of a book-buying website Nass & Lee (2000) discovered that personality traits of extraversion and introversion were observed in synthesised speech, much as they are in human speech. Furthermore, they observed evidence of *similarity-attraction effects*, in which participants displayed positive preferences and attitudes to synthesised speech personalities that matched their own. In a similar paper, Lee & Nass (2003) also observed that matching a user's personality with that of synthesised speech increased feelings of social presence, as did matching the personality of synthesised speech with textual content in contrast to when these were mismatched. However, the authors also observed stronger effects of social presence in extroverted personality conditions in contrast to introverted conditions. The similarity-attraction effect was also researched by Dahlbäck et al. (2007), in which choosing a voice that matches the accent of the user, rather than an accent related to the information being described, led people to view a vocal source as more informative and likeable, overriding any perceived expertise effects. Other research on medical IVR interactions found no effects of either user gender or voice personality conditions on user behaviours such as self-disclosure (Evans & Kortum, 2010), which suggests that manipulating certain voice characteristics may be less impactful in some interaction scenarios than others.

| Research topic | Sub-topic | | Total |
|---|---|---|---|
| System speech production | Synthesis | 8 | 18 |
| | Content | 7 | |
| | Temporal aspects | 2 | |
| | General system speech production | 1 | |
| Modality comparison | Keyboard and/or mouse | 10 | 16 |
| | Digital pen | 3 | |
| | Graphical | 2 | |
| | Gesture | 1 | |
| User speech production | General user speech production | 3 | 6 |
| | Addressee identification | 2 | |
| | Alignment | 1 | |
| Assistive Technology & Accessibility | | | 6 |
| Design Insight | | | 5 |
| Experiences with IVRs | | | 5 |
| Speech Technology for Development | | | 4 |
| IPA Experience | | | 4 |
| User Memory | | | 3 |
| Miscellaneous | | | 1 |

**Table 6. Number of papers in each research topic and sub-topic.**

Effects of voice-matching were also investigated by Truschin et al. (2014). In research using speech to read emails during driving, an interface that used synthesized voices that matched email sender characteristics led to better comprehension scores than when a single voice was used for all emails. However, this negatively impacted driving performance. Another paper assessed the impact of voice familiarity on user reactions and levels of interruption and comprehension (Bhatia & McCrickard, 2006). The authors found that while unfamiliar voices are the least interruptive, perhaps due to being filtered out more than familiar voices, quicker reactions were observed when participants heard their own voice. A further two papers examined different aspects of speech synthesis. Gong & Lai (2001) compared users' task performance and perceptions when interacting with only synthesised speech or a combination of synthesised and human speech in a virtual assistant interface. The results showed that while participants preferred the combined voice, believing they performed better on tasks in this condition, their actual performance was better when interacting in the synthesised speech only condition. Kallinen & Ravaja (2005) compared perceptions and physiological reactions towards hearing news read on a computer at a slow or fast speech rate, observing the fast rate as more arousing yet less understandable than the slow rate. Furthermore, the fast rate was judged more positively by younger participants, while older participants were more positive towards slow paced news delivery.

The second most common sub-topic in system speech production included papers that focused on the *content* of speech, as opposed to the vocal qualities in the speech synthesis research described above. These papers were interested in the message and language used by a system, including lexical choices and temporal features, and how these impacted on users. Two of these papers comparatively explored different styles of dialogue presented to users. Hu et al. (2007) compared "user-modeled summarize and refine (UMSR)" and "summarize and refine (SR)" approaches in voice-based browsing during simulated driving tasks. Despite it resulting in more minor driving errors, the authors found USMR - which presented options that best matched user preferences - was more efficient. Hofmann et al. (2014) also explored information presentation in driving simulations. Their research compared two versions of an in-car speech dialog system - one command-based and one conversational - on usability and driver distraction. Evidence suggested the command-based dialog was accepted slightly more positively than conversational dialog, possibly a consequence of the system correctly interpreting users' utterances more frequently. Walker et al. (1998) had a somewhat similar focus with a spoken language interface for email. In comparing dialogues that were either system controlled (system initiative dialogue) or flexibly controlled by the user or system (mixed initiative dialogue), the authors found that 'system-initiative dialogue' was preferred, although 'mixed-initiative dialogue' was more efficient in terms of time and number of turns to complete email tasks. Three content-focused papers also explored expectations and perceived appropriateness of system language use. Porayska-Pomsta & Mellish (2013) evaluated a model of tutorial feedback in comparison to responses by a human tutor, finding there was no significant difference between a system's best preferred output and that of a human tutor. Buchheit & Moher (1990) assessed people's expectations of a hypothetical computer partner compared to a human partner in terms of their levels of assertiveness. They found that people expect computer partners to be more assertive and less successful at being able to 'soften the impact' of speech acts. Another paper exploring studies on system politeness found that the use of different politeness strategies led people to have less positive attitudes towards an automated voice banking service (Wilkie et al., 2005). Finally, one content-focused paper assessed evaluations of a spoken dialogue system designed to tailor its use of acknowledgements in accordance with the perceived 'ephemeral emotions' of its user (Tsukahara & Ward, 2001). Results indicated that participants preferred the system that had this feature to the system than without it.

Two papers focused on *temporal aspects* of a system's speech i.e. *when* as well as *how* information should be delivered. Iqbal et al. (2011) looked at being proactive with road condition alerts to drivers engaged in phone conversations. They showed that interventions lowered collisions and turning errors compared to trials without interventions. Moreover, more descriptive messages that were explicit about upcoming conditions were perceived more positively than shorter, more general messages. Kousidis et al. (2014) also investigated dialogue systems in driving simulation tasks, when testing a situationally aware system that refrained from interruptions when higher driver attention was required. Their results showed improvements in driving performance and attention to the content of speech, compared to when situational awareness was not available, but noted similar levels of performance and attention when drivers could control the system.

Finally, one paper in the review sample was categorised as investigating *general system speech production*, as the paper demonstrated a focus on both synthesis and content produced by a system. Clark et al. (2016) examined people's responses to an interface using vague language, while also exploring the effects across different synthesised and human voices. Participants perceived vague language as more appropriate when used by a human recorded voice compared to any of two synthesised voices, with the authors suggesting synthesised speech may be less capable in executing its social functions.

*5.2.2 Modality comparison*

Sixteen papers in the review compared the use of different modalities, exploring concepts including user performance and UX. Example comparisons include comparisons of single modalities (e.g. Begany, Sa, & Yuan, 2015; Murata & Takahashi, 2002; Oviatt et al., 2008), as well as unimodal and multimodal combinations (e.g. speech and gesture (Hauptmann & McAvinney, 1993)).

The results of these modality comparisons are mixed. Ten papers explored speech, keyboard/text, and mouse input. Seven papers explore just speech and keyboard/text input. Two papers report a negative impact of using speech, with speech reducing ease of system use (Begany et al., 2015), and a user's sense of agency (Limerick et al., 2015) compared to more traditional input modalities such as keyboard and mouse. Conversely, two papers report improvements when using speech as an input modality. Greater benefits were observed among elderly compared to younger users, particularly for those not accustomed to keyboard input (Murata & Takahashi, 2002). Furthermore, with a voice-driven video learning interface, voice was used more than typing, with it being seen as more useful (Culbertson, Shen, Jung, & Andersen, 2017). Two further papers compared speech and text within a dialogue interaction. Le Bigot et al. (2006) found performance with an information retrieval system improved over time, regardless of using speech or text. However, similar experiments observed lower efficiency with such systems (Le Bigot et al., 2007). Both papers also reported effects on language use, with higher uses of pronouns in user speech input. Another paper compared speech and text as system outputs when users answered open-ended questions to determine their levels of creativity and self-disclosure, with higher levels of creativity observed in users' answers when exposed to text output (Wang & Nass, 2005).

Three papers explored the use of a mouse alongside speech, either with or without a keyboard, and observed mixed preferences in modality choice. In comparing unimodal and multimodal systems, participants were shown to prefer using a mouse for navigating within a graphical interface but using natural language for entering data (Melichar & Cenek, 2006). A different multimodal system for radiologists showed users preferring speech unless their hands were already on the keyboard or mouse, though it encountered difficulties including some users' low tolerance for speech recognition errors, as well as the difficulty of remembering some spoken commands (Lai & Vergo, 1997). Bekker et al. (1995) examined the use of only speech and mouse input with a document-annotation system, observing more errors with speech and a user preference for mouse input.

Three papers in this category also explored the use of *digital pen* input alongside other modalities. Two of these papers observe the unimodal and multimodal use of pen and speech input. Suhm et al. (2001) examined error correction for speech interfaces, finding a preference for multimodal correction. Specifically, skilled typists preferred keyboard and mouse input, but the authors predict higher performance could be generated for all users with speech and pen input instead. When speech and pen inputs were available, a shift to multimodal communication was observed when cognitive load increased (Oviatt, Coulston, & Lunsford, 2004). A further paper showed cognitive load was best managed with speech input compared to pen input with a simulated tutoring system (Oviatt et al., 2008).

Another three papers explored speech in combination with *graphics* and *gesture*. Two assessed the use of *graphical* modalities. In a mobile multimodal interaction, decreasing graphical input efficiency resulted in higher speech usage (Schaffer et al., 2015). In comparison with only graphical output, the higher the level of spoken feedback present with a timetable information system, the more that participants rated the system as human-like (Qvarfordt, Jönsson, & Dahlbäck, 2003). Finally, in comparing speech with *gesture*, as well as with a combination of the two, no significant differences were found when using them to manipulate graphical objects (Hauptmann & McAvinney, 1993), although user preferences were observed for speech and gesture combined.

*5.2.3. User speech production*

Six papers focused on *user speech production*. Of these, three were categorised as investigating *general user speech production*. Amalberti et al. (1993) showed that people adapt their language choices according to their partner models but noted that differences between human and computer speech choices decreased as people got more familiar with the interaction. People also tended to use fewer fillers (e.g. "um", "err"), request confirmation and repetition more, and use fewer topic shifts in computer compared to human interaction. Kumar, Paek & Lee (2012) compared existing dictation with "Voice Typing" - a speech interaction model that transcribes users' utterances as they are produced, allowing for error identification in real-time. In using this, their study showed a reduction in error rate and certain cognitive demands compared to dictation. Another paper explored the impact of spoken translation software on cross-lingual dialogues (Hara & Iqbal, 2015). During experiments, participants were observed adapting their speech and comprehension due to imperfections in system-produced translations, and the authors accordingly formulated a set of design guidelines for such systems.

Two papers examined *addressee identification*. In aiming to establish better models for differentiating between computer or human-directed utterances by users, Lunsford, Oviatt & Coulston (2005) observed that participants reduced the loudness of their voice to indicate self-talk, compared to higher amplitudes used for system-directed speech. Lunsford & Oviatt (2006) investigated the accuracy of people's judgements of whether a speaker's intended interlocutor is a human or computer. Participants were more accurate in identifying computer-intended interlocutors than human, particularly when presented with visual information alone, and both gaze and tone of voice were important factors in making judgements.

One paper focused on *alignment* - where speech choices of one partner in conversation are mimicked by the other. The role of *partner models,* i.e. people's perceptions of a system ability to understand the user, in this mimicry was explored in one paper (Cowan et al., 2015). The authors observed that, although people's partner models were affected by the humanness of the synthesis used, this did not significantly impact levels of syntactic alignment between human or computer partners.

*5.2.4. Assistive Technology & Accessibility*

Speech interface usability and user experience was explored in different specific communities in the reviewed papers - i.e. users with specific needs or requirements, or users interacting with a specific type of system. Six of the papers reviewed were categorised as researching *assistive technology & accessibility*. Assistive technology is generally defined as "Adapted or specially designed equipment, products and technologies that assist people in daily living" (World Health Organization, 2001, p. 174). Accessibility often includes the use of assistive technology in designing environments for improving usability with specific people (LaPlante, 1992). The majority of these papers (N=5) deployed some form of prototype system or device in their research. Alm et al. (1993) assessed the usability of an assistive dialogue prototype for users who have severe physical disabilities and do not speak. In using a system that helps these users take part in conversation, one subject was observed to increase their number of words in conversation, compared to using a more traditional word board and a speech device that stored specific phrases and names. Piper & Hollan (2008) adapt a participatory design process in creating an interactive tabletop display with audio input to assist conversations between physicians, deaf patients and interpreters. They discussed the potential for using tabletop displays for enhancing privacy and independence for deaf patients, for example in interactions with doctors or lawyers regarding personal matters. Two papers explored system prototypes designed for people with physical impairments. One of the papers assessed a speech interface with a smartphone (Corbett & Weber, 2016), finding that existing approaches in voice interaction design do not necessarily translate into mobile interaction spaces. The authors highlight the need to explore mobile-specific theories, rather than always borrowing those from desktop-based interaction. Another paper compared users' interactions with a voice-based mouse emulator with motor impaired and non-impaired individuals (Harada et al., 2009). Sato et al. (2011) conducted studies with blind participants, analysing the benefits of a voice-based web browser plugin, which uses a secondary voice to provide contextual information alongside the primary voice. When comparing this to traditional screen reader software, results showed the plugin system brought the users increased confidence, though did not improve task performance. Finally, Pak et al. (2008) also examined auditory computer interfaces, though instead of assessing a prototype system as such, looked at the impact of age-related differences in spatial ability. Their findings showed evidence of age related decline in cognitive abilities, particularly spatial, and the negative impact on the performance on the auditory task performed.

*5.2.5. Design Insight*

Five other papers also explored the use and deployment of systems in order to generate *design insight* for further iterations of development. For example, Yankelovich et al. (1995) focused on reporting the design of an experimental speech interface, *SpeechActs,* which aimed to access existing GUI based applications (e.g. email and calendar tasks). Interface testing led to a number of design guidelines including the need to develop effective conversational structures, lexicons, information flow and organization. Moran et al. (2013) investigated team responses to an agent's instructions in a pervasive game, and provided design guidelines on concepts including trust, compliance, and reliability. One paper developed a system designed for healthcare workers to interact with a voice agent for multiple functions and presented the results of its deployment in a user study (Sammon et al., 2006). Hakulinen et al. (2004) explored the design and iterative development of an integrated tutor for teaching new users to operate a speech-based email reading application. Finally, one paper had a different approach to design insight. Derriks & Willems (1998) adopted a corpus analysis approach to identify and classify negative feedback phenomena in human-machine information dialogues, with a view to future dialogue modelling.

*5.2.6. Experiences with IVRs*

Another five papers assessed *Experiences with IVRs*, exploring design choices including dialogue strategies, vocal characteristics, and methods of directing calls. Wolters et al. (2009) evaluated strategies for alleviating working memory load when scheduling healthcare appointments over the phone. Their results showed both older and younger users were more efficient in booking appointments when systems presented more options to users, and also avoided explicitly confirming choices. Furthermore, working memory span was not observed to affect user recall of appointments. Wilkie et al. (2007) demonstrated that providing hidden menu options versus explicit options to encourage proactive user requests may prevent users from successfully completing phone-based banking tasks. Litman & Pan (2002) evaluated an online train schedule retrieval system accessed via telephone, revealing performance improvements when the system predicts ASR problems and adapts to more conservative dialogue strategies. In another paper, a field study compared call routing in a cell center environment (Suhm et al., 2002). The results showed improvements in usability for natural language call routing compared to traditional touch-tone menus, as well as the potential for reducing call center costs. Another paper focused on aging and its effects on IVR experiences (Dulude, 2002). Usability ratings were found to be lower for older users than younger users, with confusing choices, speech rate and length of options accounting for the majority of problems for older users.

*5.2.7. Speech Technologies for Development*

Four papers explored the domain of *Speech Technologies for Development*. There were similarities in these papers with those focusing on modality comparisons, however, these four papers had a distinct focus on speech interface use to support rural communities, novice users, or people with low levels of literacy. Medhi et al. (2009) compared speech-based, text-based, and rich multimedia money-transfer interfaces, highlighting that while rich multimedia interfaces containing audiovisual output provided better task completion, the speed was quicker with speech-based interfaces and less assistance was required when using them. Similarly, Medhi et al. (2011) compare a text-based interface with a live operator for patients discussing symptoms with health workers over the phone, observing that the live operator was up to ten times more accurate than the text-based interface. Patel et al. (2009) compared dialled and speech input interfaces for farmers accessing agricultural information over the phone, discovering that dialled input outperformed speech in both task completion rates and perceived user difficulty. Raza et al. (2013) had a somewhat different approach, and explored the challenges of virally spreading awareness of, and training people in, speech-based services for users in communities with low levels of literacy. The authors noted that, while the system would spread to new users quickly, a rapid declining interest saw the majority of people using the system for only a few days.

*5.2.8. IPA Experience*

Four papers explored people's experiences with current IPAs. Leahu et al. (2013) discuss the flexible, rather than static, nature of "human" and "machine" identity categories in this context, and possible experimental design implications. Luger & Sellen (2016) describe power users' experiences with IPAs, highlighting a dissonance between people's mental models and the reality of what an IPA can do. They also found that people lack trust in IPAs' ability to conduct tasks effectively, and that users saw speech interaction as something that had to be learned. Porcheron et al. (2017) report how IPAs are used by multiple users at once, mapping the collaborative mechanism and structure of the interaction, as well as highlighting the challenges in mutual social silences. Finally, Cowan et al.

(2017) examined barriers to the more frequent use of IPAs, including usability difficulties, social embarrassment with public use, and the negative impact of human likeness on UX.

*5.2.9. User Memory*

Three papers assessed the effects of interface design on *user memory*. User recall of menu options in IVR systems was significantly impaired when five or more options were presented, and with suffixes impairing memorization regardless of message (Le Bigot, Caroux, Ros, Lacroix, & Botherel, 2013). A study on cognitive load for speech interface users scheduling health appointments showed that working memory span did not affect appointment recall, nor did strategies used to reduce working memory such as reducing the number of menu options or providing confirmation of the system interactions (Wolters et al., 2009). Another study reported that more content was recalled when information was provided by a human speaker rather than a machine (Knutsen et al., 2017).

*5.2.10. Miscellaneous*

One paper could not be categorised. Ramanarayanan et al. (2017) tested the efficacy of crowd source evaluations of engagement levels between non-native English speakers and a computer-assisted language learning (CALL) system. While findings showed consistent ratings of engagement when the human, computer, or both were communicating to one another in the videos, there were low correlation levels between self-evaluations and third-party evaluations of engagement.

## 6. Discussion

Our review reveals the state of empirical speech interface research within the field of HCI at present. We show that much of the research tends to be published in conferences rather than journals in the field. Methodologically, papers tend to explore users' behaviours and attitudes towards a range of interactive prototypes, commercial, or Wizard of Oz systems. Although behavioural measures such as lexis and syntax and system usage were used in the research, self-report questionnaires measuring concepts like usability and user attitudes were most common. Yet many of these lack reliability or validity testing and are not consistently used across the research reviewed creating issues for validity of measurement. Most studies tend to compare aspects of a design or investigate theory inspired research questions (*usability/theory-based research*). Few papers in the review were exploratory or design based in nature. Our categorisation of research topics found that speech interface work published in HCI venues coalesced around nine main topics: *system speech production, modality comparison, user speech production, assistive technology & accessibility, design insight, experiences with IVR systems, using speech technology for development, people's experiences with IPAs, and how user memory affects speech interface interaction*. Based on these findings we point to a number of key challenges in the research, methodological approaches, and evaluation of speech interfaces to be addressed by future research efforts.

### 6.1. Research Challenges

*6.1.1. Developing theories of speech interface interaction*

Our review highlights that speech interface work has been successful in producing theories and effects that have been widely adopted by the field (e.g. the *similarity-attraction effect* (Dahlbäck et al., 2007; Nass & Lee, 2000); *consistency-attraction* (Lee & Nass, 2003)). These concepts focus on explaining user attitudes, allowing researchers to predict, hypothesise and interpret the impact of particular design decisions. Yet our review showed a similar set of concepts or theories for understanding user language choices in speech interface interactions is lacking in the literature. This insight is critical as understanding what influences users' language input is as fundamental to speech HCI as understanding touch or selection-based behaviours in other interaction modalities. Although few formal effects or theories have been explored in the literature, our review does illuminate a consensus and emerging debate that could be used as a foundation for theory building in this regard. A number of papers reviewed propose that language in speech interface interaction is driven by the assumptions users have of partner abilities (i.e. our partner models) and that they adapt speech choices accordingly (e.g. Amalberti et al., 1993), similar to mechanisms proposed in human-human communication (Branigan, Pickering, Pearson, McLean, & Brown, 2011; Brennan & Clark, 1996). The evidence for this in our review is mostly based on the differences in speech production between human-human and human-computer dialogue, although a study that looks at the impact of system design debates the influence of partner models on all language choices (Cowan et al., 2015). As well as influencing user language choices, these expectations may also affect evaluation of system output. One paper demonstrated the relationship between an interface's voice and the perceptions of politeness in receiving computer instructions, based on expectations of linguistic capabilities (Clark et al., 2016). A key debate around the role of partner models seems to be emerging in HCI based publications, which reflects debates on language production more generally (Brennan & Clark, 1996; Horton & Keysar, 1996; Keysar, Barr, & Horton, 1998). An opportunity for the HCI community lies in building on this emerging consensus. From this a more robust theory of language production and interlocutor perceptions may grow, which will be able to inform design and interaction science research (Howes et al., 2014) in the speech domain.

*6.1.2. Achieving critical mass*

The review also shows the fragmented nature of research within the area, highlighting a need for developing critical mass in a number of these topics. For instance, within the nine primary topics across the work, the majority held less than seven research papers. Even within topics the work tends to be fractured into specific sub-topics with few papers representing these, making it hard to identify a consensus or a summary of results. For example, system speech production research (Section 5.2) explores four distinct sub-topics; speech synthesis (e.g. Nass & Lee, 2000), content (e.g. Hoffmann et al., 2014), temporal aspects (e.g. Iqbal et al., 2011), and general speech production (Clark et al., 2016), with each study varying in the concepts explored within those sub-topics. Similar variation was observed for papers comparing speech with other modalities. While diversity across speech HCI research is encouraging, themes discussed in this paper point to a noticeable fragmentation of topics explored. However, this was not the case for all areas. Four papers discussing speech technologies for development (Section 5.2.7), for example, tended to hold a common thread of exploring interface use with novice users and people with lower levels of literacy. They also provide example benefits and drawbacks of speech interface design and use with these communities (e.g. Medhi et al., 2009; Raza et al., 2013). The growing interest in speech interfaces should be used to contribute greater cohesion to the current body of knowledge in speech-based HCI. Indeed, increased focus could help develop and embed new and existing theories, concepts and paradigms (e.g. those in Section 6.1.1.) across the numerous research topics within the area.

*6.1.3. Speech interaction design work is required*

Our review showed that one area that is perhaps most evidently lacking in the HCI domain is design related research. We found few design papers in the ones reviewed (e.g. Corbett & Weber, 2016; Yankelovich et al., 1995). Approaches commonly used in interaction design research such as cultural probes (DeHaemer & Wallace, 1992), contextual inquiry (Holtzblatt & Jones, 1993), and participatory design more generally (Schuler & Namioka, 1993) tended not to be used in the research we reviewed. A number of papers took more of a usability engineering-based approach, comparing interface designs e.g. menu structures (e.g. Wilkie et al., 2005; Wilke et al., 2007), and observing their effect on user attitudes and performance. Although some of the design papers in the review mention specific considerations for speech interface design, the lack of design research means that there are currently no clear design considerations or robust heuristics for developing user centred speech interactions, although texts exist to support the design of voice user interfaces (VUIs) (e.g. Pearl, 2016). With the popularity of speech interfaces increasing, there is a real need for this type of knowledge to guide improvements and development of future speech interfaces. We also need to identify what design methods and theories work best in a speech context, rather than solely importing those that have worked on other modalities and contexts without critical reflection (Corbett & Weber, 2016). This can be achieved if the design community in HCI embraces speech interfaces more widely in the future.

*6.1.4. Investigating multiple user contexts*

The majority of papers in the review researched single user interactions. Yet with IPAs like Alexa or Google Home being used in social spaces, the interaction opportunities and challenges of multiple user scenarios need to be further understood. Research in the HCI community has begun to shed light on group-based IPA interactions, in particular the dynamics of user behaviour in this experience (e.g. Porcheron et al., 2017). The technical challenges in terms of recognizing who is speaking (speaker diarization) and whether they are addressing the system are well researched (Batliner, Hacker, & Nöth, 2008). The HCI community is well placed to inform these technical advancements, particularly by understanding users' expectations, behaviours, social dynamics and the interaction barriers that may limit the effectiveness of speech interfaces in group situations. The study of situations that include multiple users and multiple speech interfaces would be highly valuable as this is likely to become more common in the near future.

**6.2. Methodological & Evaluation Challenges**

*6.2.1. Improving measure reliability, validity and consistency*

Across the papers in our review, there was a range of objective and subjective concepts being measured, including task performance (e.g. Oviatt et al., 2008), production of lexis and syntax (e.g. Cowan et al., 2015), system usage (e.g. Schaffer et al., 2015), and user recall (e.g. Knutsen et al., 2017). A high proportion of research in our review used self-report questionnaires to measure other concepts like user satisfaction, usability, user attitudes towards speech interfaces and general user experience. A number of these self-report measures lacked any reliability or validity testing. Self-report questionnaires specific to speech interfaces that have been assessed for internal reliability do exist in the literature (e.g. SASSI (Hone & Graham, 2000)), although these need further validity testing (Hone, 2014; Hone & Graham, 2000). We were surprised to see that, although SASSI was deployed once in the reviewed papers (Hofmann et al., 2014), the scale has rarely been used in HCI publications. The field therefore needs to make a concerted effort in developing well-validated, reliable subjective measures for more UX related dimensions in speech interaction. This is made more challenging as our review shows a lack of consistency in the concepts being measured across papers, which presents difficulties for the domain in building robust measures and a body of knowledge around specific concepts or paradigms. Discussion within the HCI field is needed to map out the measures, concepts and paradigms that are pertinent to this research area, so as to identify the priority areas for efforts of metrics development to focus on. This type of mapping could also bring more cohesion to speech HCI work as a whole.

*6.2.2. Evaluating speech interfaces in real world contexts*

Throughout the papers in the review, there was a strong tendency to use lab-based approaches to investigate speech interface interactions. These lab experiments often use controlled prototypes and/or Wizard of Oz simulated systems. This work is essential in developing the necessary science in the field. Yet further effort is also needed in understanding how these lab-based findings transfer to real-world contexts, as well as comparing interactions across multiple contexts. As Corbett & Weber (2016) highlight, theories may not automatically translate from one context of interaction to another. More studies using *in the wild* experimental approaches would be useful in this regard. Indeed, we feel that more qualitative work to explore wider issues in speech interface interaction should be encouraged. This type of work creates valuable insight into the impact of context on speech interface use, whilst also shedding light on what types of situations these interfaces are used in and what types of UX issues emerge. Our review indicates that there has been some recent progress towards this (e.g. Luger & Sellen, 2016; Porcheron et al., 2017; Cowan et al., 2017). Future work should look to build on these efforts.

*6.2.3. Reducing barriers to building speech interfaces*

As mentioned above we found a lack of design related work on speech interfaces. The scarcity of design research may speak to a perceived high barrier in developing operational speech interface prototypes to explore in these types of studies. A number of packages and toolkits do exist that can help in developing prototypical speech interfaces such as OpenDial (Lison & Kennington, 2016), IrisTK ([www.iristk.net](www.iristk.net)) or Aspect Prophecy ([www.aspect.com/](www.aspect.com/)). Amazon's Skills Kit ([https://developer.amazon.com/alexa-skills-kit](https://developer.amazon.com/alexa-skills-kit)) also allows people to develop skills for Alexa that could be deployed as prototypes for interaction studies. Highly flexible and easy to use tools to develop speech interfaces for research and prototyping purposes should be further encouraged to facilitate speech interface prototyping and design.

**6.3. Limitations**

The aim of this research was to identify the main trends, themes and methods used in speech research published specifically in core HCI venues. To do this we focused on a comprehensive list of the top HCI publication venues, based on Google Scholar, Thomson Reuters and Scimago rankings. This allows us to focus specifically on HCI related papers. However, there are a number of studies published in other fields that are relevant to speech HCI work, including cognitive psychology, ergonomics, linguistics, and speech technology (e.g. Branigan et al., 2011; Large et al., 2017; Cowan & Branigan, 2015; Mendelson & Aylett, 2017). Although these areas are beyond the scope of the review, similar topics are present in papers published in these areas. Future reviews should look to concentrate on these fields and compare directly with the findings from this research. Our research also focused on empirical studies, and thus omitted research without user evaluation or user-based data collection that may have been influential to the field. We also decided to exclude research on embodied conversational agents and robotics in speech interfaces, because of the danger of embodiment-based factors confounding speech-related effects. Future reviews should focus on these domains.

## 7. Conclusion

This paper maps out the trends and findings of speech research published in HCI, in the hope of stimulating further speech-based research by describing the current state of the field. Based on the review of 68 papers from core HCI publication venues, we highlight nine primary research topics, and identify the key methodological approaches taken in the research reviewed. From this we identify key research, methodological, and evaluation challenges in developing theory and design work in this area, expanding contexts of human-system evaluation, reducing technical barriers to research, and improving consistency and rigor in evaluation measures.